\documentclass[11pt]{article}

\usepackage[preprint]{acl}


\usepackage{times}                 % 正文用 Times 字体（ACL 要求）
\usepackage{latexsym}              % 一些数学符号
\usepackage[T1]{fontenc}           % 让带重音的字母能正确断行
\usepackage[utf8]{inputenc}        % 支持 UTF-8 输入编码
\usepackage{microtype}             % 微调字间距，排版更美观、更省空间
\usepackage{inconsolata}           % 美化等宽（代码）字体

\usepackage{graphicx}              % 插图必备：\includegraphics
\usepackage{amsmath,amssymb}       % 公式必备：align 环境、各种数学符号
\usepackage{booktabs}              % 画"三线表"必备：\toprule \midrule \bottomrule
\usepackage{float}                 % 提供 [H] 选项，强制表格/图片固定在当前位置
\usepackage{makecell}             % 表头单元格内换行（Table 1 用）
\usepackage{placeins}

\title{Agentopia on a Consumer GPU:\\ A Reduced-Scale Long-Horizon Port with an 8B Model}

\hypersetup{
  pdftitle={Agentopia on a Consumer GPU: A Reduced-Scale Long-Horizon Port with an 8B Model},
  pdfauthor={Luo Huan}
}

\author{Luo Huan \\
  Shenzhen University \\
  Shenzhen, China \\
}

\begin{document}
\maketitle

\begin{abstract}
Large language model (LLM)-based multi-agent social simulation has demonstrated compelling results, but Agentopia was evaluated with 100 agents over 10 simulated years using Qwen3.5-397B-A17B, leaving the behavior of reduced-scale deployments on consumer hardware unclear.
In this paper, we implement and evaluate a reduced-scale Agentopia port on a single NVIDIA RTX 5070 Ti (12\,GB VRAM) using Qwen3-8B-AWQ, a 4-bit quantized model.
We introduce three structural adaptations for this setting: (1)~system-managed layered memory compression, (2)~four activity blocks per simulated day, and (3)~explicit physical- and mental-health state variables.
Across three independent stochastic runs, two runs completed 52 weeks and the third completed 50 weeks before reaching the context limit, totaling 154 system-weeks (770 agent-weeks).
No agent died, and no threshold-based health warning was logged; activity records containing at least one \texttt{NO\_RESPONSE} field occurred at rates of 10.15--10.29\% across runs.
A 52-week memory-off run tied L2/L3 artifact production to layered memory; a separate 10-week comparison associated four daily time blocks with $2.72\times$ more finalized records and lower lexical duplication, but a higher missing-field rate.
These comparisons do not support causal behavioral claims.
We release validated configurations, derived audits, analysis scripts, aggregate figure data, and our implementation changes in a public fork; raw runs and initial persona data are excluded because their redistribution provenance is not fully resolved.
\end{abstract}

\section{Introduction}

Large language model (LLM)-based agents have demonstrated remarkable capabilities in individual tasks, but their behavior in long-term multi-agent social simulations remains underexplored.
Agentopia \citep{wang2026agentopia} recently introduced a framework where 100 LLM-based agents live, interact, and learn in a shared society over 10 simulated years.
Its primary experiments use Qwen3.5-397B-A17B and substantially more compute than a consumer laptop.
This leaves a practical question unanswered: what scale and duration of Agentopia-style simulation are attainable on hardware accessible to individual researchers?

In this paper, we study a reduced-scale port with five active agents, one apartment world, and a target horizon of one simulated year.
We deploy Agentopia on an NVIDIA RTX 5070 Ti (12\,GB VRAM) using Qwen3-8B-AWQ, a 4-bit quantized model served via vLLM \citep{kwon2023vllm}.
To operate within this setting, we add three structural adaptations: (1)~system-managed layered memory compression, (2)~four activity blocks per simulated day, and (3)~explicit physical- and mental-health state variables.
Across three independent stochastic runs, the system completed 52, 52, and 50 weeks, respectively.

The two full runs reached one year, while the third exposed a context-budget boundary at week~51; activity-record-level \texttt{NO\_RESPONSE} rates were 10.15--10.29\%.
Disabling layered memory removed L2/L3 artifacts, establishing an implementation dependency but not a downstream behavioral effect.
In a 10-week comparison, the archived M+T run contained $2.72\times$ as many finalized records per agent-week as the archived M run and showed less lexical reuse but more missing fields; the unseeded, one-run-per-condition design prevents causal claims.
No deaths or threshold-based health warnings occurred, but incomplete health ablations prevent attribution to that module.

We position this work as an empirical account of a reduced-scale, long-horizon agent society on a consumer GPU, not as a same-scale reproduction of the original Agentopia experiments.

\section{Related Work}

\paragraph{LLM social simulation systems.}
Generative Agents established a widely used architecture that combines memory, reflection, and planning in an interactive sandbox \citep{park2023generative}.
Subsequent systems have emphasized complementary goals: SOTOPIA formalizes interactive evaluation of social intelligence \citep{zhou2024sotopia}; Agentopia targets 100-agent, multi-year life simulation \citep{wang2026agentopia}; and AgentSociety and GenSim focus on parallelized or general-purpose simulation infrastructure \citep{zhang2025agentsociety,tang2025gensim}.
Rather than competing on population scale, our study asks what longitudinal evidence and failure boundaries can be obtained from an Agentopia-style system on a single 12\,GB consumer GPU.

\paragraph{Validity and long-horizon evaluation.}
LLMs can reproduce selected behavioral-study outcomes or conditional response distributions \citep{aher2023simulate,argyle2023outofone}, but apparent success can depend strongly on information access and evaluation design \citep{zhou2024reallife}.
Surveys and methodological critiques therefore call for calibration, robustness checks, reproducible protocols, and explicit limits on population-level inference \citep{gao2024survey,zeng2026toohuman}.
Recent SocialLLM work similarly audits multi-turn trajectories and separates autonomous action selection from empirical calibration \citep{star2026auditing,lazzaroni2026calibrated}.
We follow this evidence-bounded view by reporting completed horizons, artifact coverage, lexical reuse, and termination conditions without treating five agents as a representative population.

\paragraph{Memory and resource-constrained deployment.}
Long-horizon evaluations expose history-dependent challenges across repeated interactions \citep{goel2025lifelong}, while memory-management studies show that how experiences are retained and consolidated can affect later experience-following behavior and computational cost \citep{xiong2026memory,zhang2026lightmem}.
Our system-managed L2/L3 pipeline is motivated by this problem, but we evaluate artifact production rather than claiming learning or a causal behavioral benefit.
At the deployment layer, AWQ provides weight-only quantization for memory-constrained inference \citep{lin2024awq}, and PagedAttention supports memory-efficient LLM serving \citep{kwon2023vllm}; these are components of our stack, not contributions re-evaluated by this study.

\section{System}

\begin{figure*}[t]
\centering
\includegraphics[width=\textwidth]{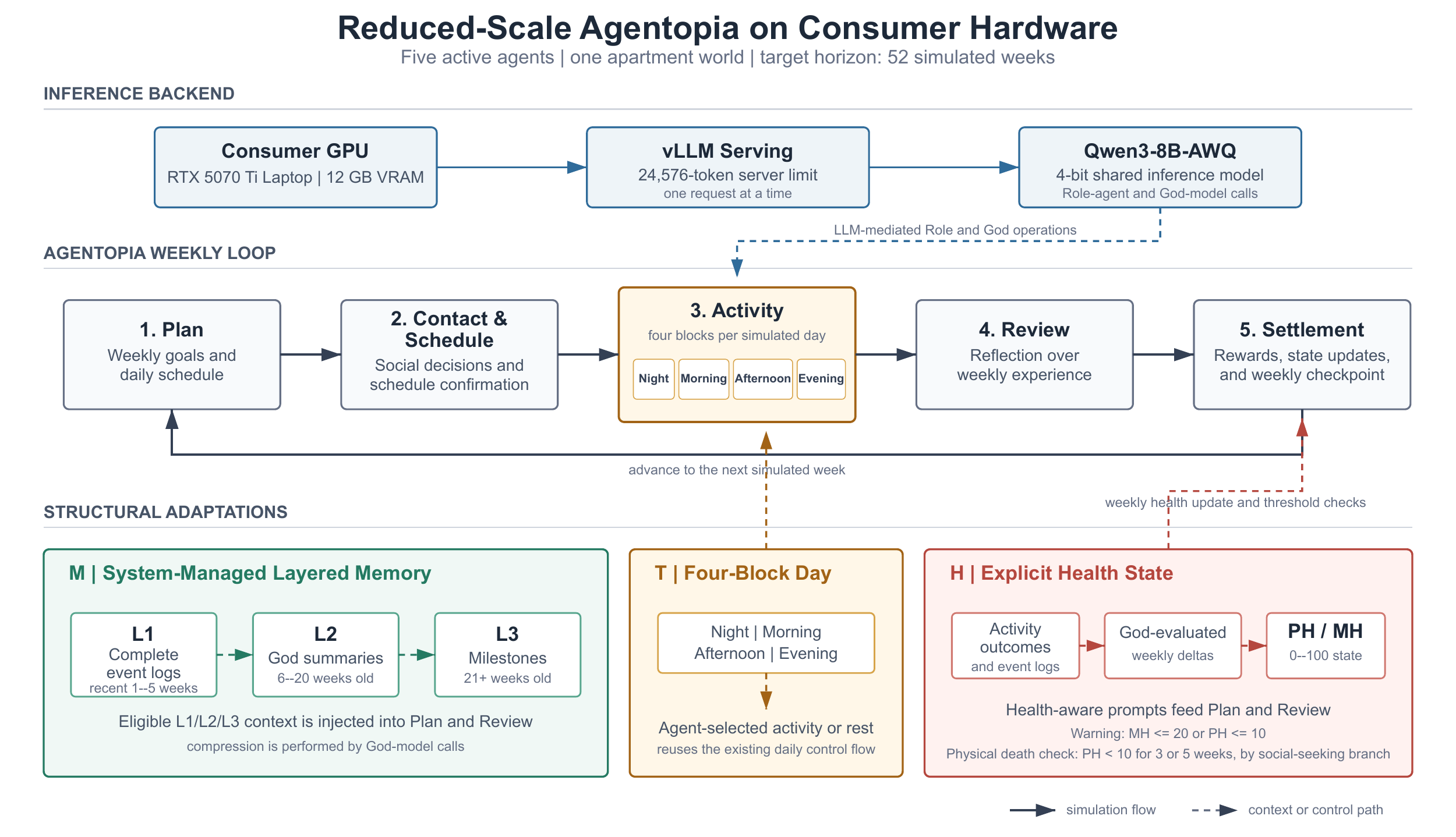}
\caption{Architecture of the reduced-scale Agentopia port. A shared Qwen3-8B-AWQ endpoint served by vLLM executes Role-agent and God-model calls on a 12\,GB consumer GPU. The weekly loop retains planning, contact and scheduling, activity, review, and settlement, while the M/T/H adaptations add system-managed layered memory, four activity blocks per simulated day, and explicit physical- and mental-health state. Solid arrows denote simulation flow; dashed arrows denote context or control paths.}
\label{fig:architecture}
\end{figure*}

Figure~\ref{fig:architecture} summarizes the inference stack, weekly simulation loop, and the three structural adaptations used in our reduced-scale port.

\subsection{Environment and Baseline}

We adopt the Agentopia framework \citep{wang2026agentopia} with five active agents cohabiting an apartment world for up to 52 simulated weeks.
Our implementation retains the framework's plan, contact and scheduling, activity, review, and settlement operations; rewards are computed during settlement according to the configured period.

Our hardware setup consists of an NVIDIA RTX 5070 Ti (12\,GB VRAM) running Ubuntu 22.04 under WSL2.
We serve Qwen3-8B-AWQ \citep{yang2025qwen3,lin2024awq}, a 4-bit quantized 8-billion-parameter model, via vLLM \citep{kwon2023vllm} with a server-side context limit of 24,576 tokens, role model output capped at 3,584 tokens, God model output at 3,840 tokens, temperature 0.7, and concurrency limited to one request (\texttt{max\_num\_seqs=1}) to avoid memory pressure.
The archived run configuration records a client-side context value of 28,672 tokens; the effective limit was the lower 24,576-token server boundary.
We progressively scaled the simulation through a ladder of tests from T1 (1 agent $\times$ 4 weeks) to T6 (5 agents with a target of 52 weeks).
The three T6 runs were independent stochastic restarts, but no global run-level seed was fixed; we therefore do not describe them as controlled-seed replicates.

\subsection{Layered Memory Compression}

The original Agentopia provides recent weekly diaries together with persistent scratchpad files that agents manage themselves.
Detailed diaries outside the recent window leave the prompt unless an agent has promoted salient information into a scratchpad, so long-term retention depends on agent-initiated memory management rather than a systematic compression schedule.

We add a system-managed three-layer memory architecture with progressively coarser representations.
\textbf{L1} stores complete event logs for the five most recent weeks; agents do not generate these records themselves.
\textbf{L2} holds $\sim$200-word summaries for intermediate history, produced by the God model from L1 logs.
\textbf{L3} distills older blocks of L2 summaries into $\sim$50-word milestone entries.
When composing plan and reflection prompts, agents receive the currently eligible context from all three layers, providing a system-managed retention path for information that would otherwise depend on scratchpad updates.

In short preliminary runs with fixed review prompts, we observed that Reflection paragraphs froze into verbatim repetition after 3--5 weeks.
After the review question set was randomized, the repetitions were no longer observed in subsequent short tests, suggesting that prompt structure may have contributed to the failure mode.
The long-run experiments below evaluate whether the layered pipeline continues to produce nonempty, qualitatively varied memory artifacts; they do not isolate its effect on downstream behavior.

\subsection{24-Hour Time Granularity}

The original framework uses weekly plans with one solo-activity slot per day, and the same planned activity may recur across weekdays.
This design limits the representation of within-day rhythms and activity variation.

We divide each day into four time blocks---Night (00:00--06:00), Morning (06:00--12:00), Afternoon (12:00--18:00), and Evening (18:00--24:00).
Agents plan an activity or rest choice for each block, allowing different behavior within the same day.
The adaptation reuses the existing weekly and daily control flow while iterating over four solo-activity blocks within each day.
When disabled (via the ablation flag \texttt{time.enable\_24h}), the system reverts to the original single-solo-activity-per-day behavior.

\subsection{Health System}

The original Agentopia models vitality and four fulfillment dimensions, including mood, but it does not maintain explicit physical- and mental-health trajectories or warning and death thresholds.

We introduce two separate health indicators: \textbf{physical\_health} (0--100, initially 80) and \textbf{mental\_health} (0--100, initially 70).
Physical health is affected by sleep duration, nutrition, and illness events; mental health responds to social interactions---positive encounters (gratitude, recognition) increase it, while conflicts, rejection, and loss decrease it.
The God model evaluates health deltas each week based on event logs and activity outcomes.
Current scores are exposed to subsequent plan and review prompts through score-dependent health-awareness text.
At each weekly checkpoint, a mental-health score at or below 20 or a physical-health score at or below 10 is logged as a threshold-based health warning.
Four consecutive weeks below a mental-health score of 20 additionally produce a severe-depression-risk warning but do not force death.
For physical-health scores below 10, agents with low social-seeking are marked deceased after three consecutive weeks, whereas high social-seeking delays this outcome until five consecutive weeks.
The profile-derived social-seeking score is the quantitative extraversion trait when available and otherwise $100-\text{introversion}$; scores of 60 or higher use the high-social-seeking branch.

To reduce template-like health evaluations, one prompt cue is sampled each week from a pool of 48 emotional cues.
The health system is controlled via the ablation flag \texttt{health.enable}; its trajectories are stored separately from the original fulfillment-based subjective reward computation.

\section{Results}

We evaluate the full system (all three adaptations enabled, denoted M+T+H) across three independent stochastic restarts lasting 52, 52, and 50 weeks, respectively.
All five selected agents are active participants, yielding 154 completed system-weeks and 770 agent-weeks.

\subsection{System Stability}

The first two runs reached the 52-week target.
The third completed 50 weeks but stopped when the week-51 prompt reached the configured context boundary.
Before termination, Run~3 wrote 16 finalized activity records for the incomplete week~51.
We include these records in activity-record-level metrics because they are present in \texttt{activity.jsonl}, but exclude week~51 from completed-week totals and weekly-outcome analyses.
No infrastructure-level termination occurred within a completed week; missing activity fields are quantified separately below.
The weekly pipeline executed fully in every completed week.
These results support feasibility at the tested five-agent scale and identify context growth as a long-horizon limitation observed in Run~3.

\subsection{Observed Health Trajectories}

\begin{figure*}[!t]
\centering
\includegraphics[width=\textwidth]{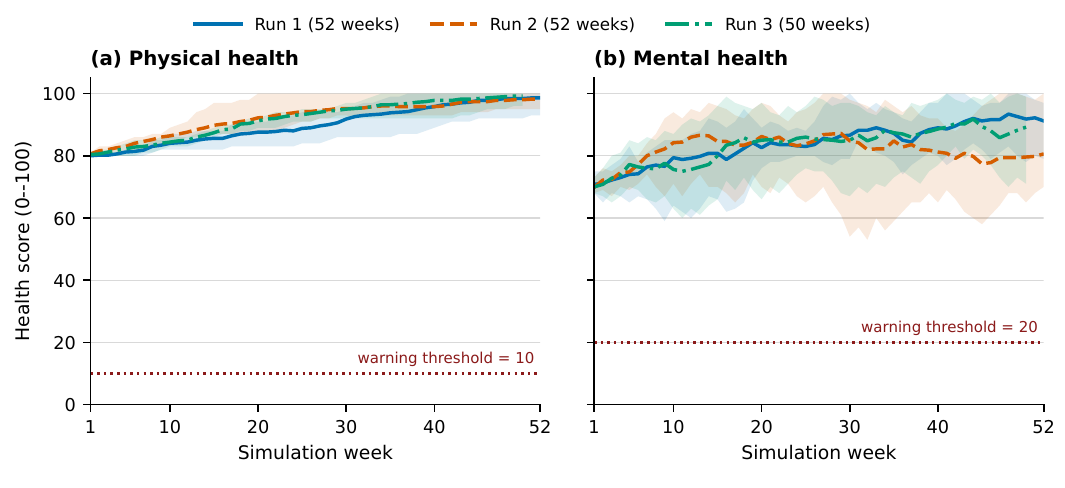}
\caption{Weekly physical- and mental-health trajectories across the three full-system runs. Lines show the weekly mean over the five active agents, and shaded bands span the agent-level minimum and maximum. Dotted horizontal lines mark the physical-health warning threshold of 10 and mental-health warning threshold of 20. Run~3 ends after week~50.}
\label{fig:health-trajectories}
\end{figure*}

Figure~\ref{fig:health-trajectories} shows the weekly physical- and mental-health trajectories, and Table~\ref{tab:health} reports the scores at each run's final checkpoint.
Across all three runs, \textbf{no agent died and no threshold-based health warning was logged}.
Physical health remained at or above 93 for all agents at the final checkpoint; the lowest mental-health value recorded across all weeks was 53 (see Table~\ref{tab:summary} for per-run minima), above the mental-health warning threshold of 20.
Because the health-disabled runs are incomplete, these observations characterize the enabled system but do not establish that the health module caused the absence of warnings.

\begin{table}[!ht]
\centering
\caption{Final-checkpoint physical-health (PH) and mental-health (MH) scores. Both range from 0 to 100. Runs~1--2 end at week~52 and Run~3 at week~50; each cell reports PH/MH.}
\label{tab:health}
{\small\setlength{\tabcolsep}{4pt}
\begin{tabular}{@{}lccc@{}}
\toprule
Agent & \makecell[c]{Run 1\\PH/MH} & \makecell[c]{Run 2\\PH/MH} & \makecell[c]{Run 3\\PH/MH} \\
\midrule
Whitfield & 100/93 & 95/70  & 96/71  \\
Morales   & 100/97 & 100/79 & 100/89 \\
Vale      & 93/92  & 96/77  & 100/96 \\
Voss      & 100/79 & 100/77 & 100/99 \\
Vieri     & 100/95 & 100/100 & 100/91 \\
\bottomrule
\end{tabular}}
\end{table}

% Table 2 — NO_RESPONSE 采用 activity.jsonl 原子记录统计，避免历史 prompt 重复计数
\begin{table}[!ht]
\centering
\caption{Activity-record-level \texttt{NO\_RESPONSE} incidence. Total records reports the number of finalized activity records, and Flagged records reports the number containing at least one \texttt{NO\_RESPONSE} field; each record is counted at most once.}
\label{tab:noresponse}
{\small\setlength{\tabcolsep}{4pt}
\begin{tabular}{@{}lrrr@{}}
\toprule
Run & Total records & Flagged records & Rate \\
\midrule
Run 1 & 4,412 & 448 & 10.15\% \\
Run 2 & 4,436 & 451 & 10.17\% \\
Run 3 & 4,236 & 436 & 10.29\% \\
\midrule
Pooled & 13,084 & 1,335 & 10.20\% \\
\bottomrule
\end{tabular}}
\end{table}

We measure missing activity fields from the finalized \texttt{activity.jsonl} records rather than counting string occurrences in generation transcripts, where the same historical marker can be copied into many later prompts.
The pooled activity-record rate in Table~\ref{tab:noresponse} is 10.20\%, and the sample coefficient of variation across the three unrounded run-level rates is 0.75\%.
This is an activity-record metric, not a call-level failure rate, because one activity can contain multiple LLM calls.

To audit output reuse separately from literal \texttt{NO\_RESPONSE} markers, we examined solo records without that activity-level flag.
Within each run and agent, normalized primary-text exact-duplicate rates were 14.48\%, 17.18\%, and 17.52\% for Runs~1--3; comparison against the preceding eight weeks using 5-token shingle Jaccard $\geq 0.80$ yielded additional near-duplicate rates of 5.72\%, 4.66\%, and 4.40\%.
Normalization applied Unicode NFKC normalization, lowercasing, and punctuation removal, and comparisons were restricted to the same activity type.
These lexical screens expose output reuse but do not measure semantic or behavioral diversity.

\subsection{Memory Artifacts and a Short Time Comparison}

Across the stochastic runs, Aaron Whitfield's week-40 L2 entries emphasized, respectively, deadline-driven family guilt, avoidance through writing, and peer conversations prompting a promised visit.
In qualitative inspection, these entries appeared non-template-like and contextually grounded.
Each active agent has 47, 47, and 45 L2 entries in Runs~1--3, respectively (Run~3 produces two fewer because it is two weeks shorter), for 695 L2 and 90 L3 artifacts in total.
Given the configured five-week recent window, all 15 agent-run archives contained an L2 entry for every expected older week; this establishes continuous artifact production, not successful recall or downstream use.
These examples illustrate that the stored summaries preserve run-specific histories; they do not establish an effect on downstream behavioral diversity.

We also audited two archived 10-week, five-agent conditions: layered memory alone (M) and memory plus four daily time blocks (M+T), with health disabled.
Their saved configurations differ only in \texttt{world.time.enable\_24h} and output directory; Table~\ref{tab:time} reports activity volume, missing fields, and lexical reuse.

\begin{table}[!ht]
\centering
\caption{Exploratory 10-week time toggle (one run per condition; health disabled). Rec./A-W is finalized records per agent-week; NR is activity-record-level \texttt{NO\_RESPONSE}. Exact/near are lexical duplicate rates in unflagged solo primary texts.}
\label{tab:time}
{\small\setlength{\tabcolsep}{3pt}
\begin{tabular}{@{}lrrrrr@{}}
\toprule
Condition & Weeks & Rec./A-W & NR & Exact & Near \\
\midrule
M (T off) & 10 & 6.36 & 1.57\% & 57.60\% & 16.40\% \\
M+T       & 10 & 17.28 & 7.75\% & 10.06\% & 6.45\% \\
\bottomrule
\end{tabular}}
\end{table}

M+T contains $2.72\times$ as many records per agent-week as M, reflecting its larger activity budget rather than higher quality; all five agents have lower exact-duplicate and higher NR rates.
With one unseeded, unpaired run per condition and different activity opportunity counts, these results are descriptive and do not establish improved semantic or behavioral diversity.

\begin{table*}[!t]
\centering
\small
\caption{Run-level summary. NR is the activity-record-level \texttt{NO\_RESPONSE} rate, D/W denotes deaths/threshold-based health warnings, and health minima cover all completed weeks. L2 entries/agent is the final-checkpoint count per active agent. Aggregate weeks are system-weeks, and aggregate NR is pooled over records.}
\label{tab:summary}
\begin{tabular}{@{}lrlrrrrcc@{}}
\toprule
Run & Weeks & End condition & Activity records & NR rate & Min. PH & Min. MH & D/W & L2 entries/agent \\
\midrule
Run 1 & 52 & Target reached   & 4,412 & 10.15\% & 80 & 59 & 0/0 & 47 \\
Run 2 & 52 & Target reached   & 4,436 & 10.17\% & 80 & 53 & 0/0 & 47 \\
Run 3 & 50 & Context boundary & 4,236 & 10.29\% & 80 & 60 & 0/0 & 45 \\
\midrule
Aggregate & 154 & -- & 13,084 & 10.20\% & 80 & 53 & 0/0 & -- \\
\bottomrule
\end{tabular}
\end{table*}

% Table 5 — ablation，前移到 summary 之后，与 summary 叠放在第 4 页顶
\begin{table*}[!t]
\centering
\caption{Completion status and observed outcomes for the full-system runs, ablation conditions, and 4-week all-off baseline. M, T, and H denote layered memory, four-block daily time, and health.}
\label{tab:ablation}
\footnotesize
\setlength{\tabcolsep}{4pt}
\begin{tabular}{@{}p{2.7cm}ccp{3.4cm}p{5.3cm}@{}}
\toprule
Configuration & Runs & Weeks completed & Status or termination & Observed outcome \\
\midrule
Full (M+T+H)  & 3 & 52, 52, 50 & Planned horizon completed (2); context boundary (1) & Zero deaths/warnings; L2/L3 artifacts present \\
$-$Memory (T+H) & 1 & 52 & Planned 52-week horizon completed & L2/L3 artifacts absent by configuration \\
$-$Health (M+T) & 2 & 21, 38 & Incomplete: week 21, resume-path slowdown; week 38, WSL I/O failure & No complete 52-week health comparison available \\
Time toggle (M vs. M+T) & 1 each & 10, 10 & 10-week comparison complete; 52-week M+H not run & $2.72\times$ records/A-W; lower reuse, higher NR; descriptive only \\
All off & 1 & 4 & Planned 4-week baseline completed & Longitudinal comparison unsupported; health warning/death mechanisms disabled \\
\bottomrule
\end{tabular}
\end{table*}

\subsection{Cross-Run Summary}

Table~\ref{tab:summary} aligns the main run-level metrics with each run's end condition.
Across runs, minimum PH was 80 in all cases, minimum MH ranged from 53 to 60, and activity-record-level \texttt{NO\_RESPONSE} rates ranged from 10.15\% to 10.29\%; Run~3 ended two weeks earlier at the context boundary.
Table~\ref{tab:ablation} summarizes coverage and completion status for the configurations discussed below.

\FloatBarrier

\section{Discussion}

\subsection{Ablation Analysis}

We completed one 52-week memory ablation, one 10-week time-toggle comparison, and two incomplete attempts to disable the health module; we did not complete a 52-week M+H leave-one-out comparison for the time adaptation.
The memory-off run (T+H enabled) completed 52~weeks with zero deaths and zero threshold-based health warnings, showing that this configuration remained operational without layered memory. Physical health ranged from 98 to 100 across agents at the final checkpoint, and mental health from 74 to 98. As expected from the configuration, its L2 and L3 directories were empty and no compressed summaries were available to agent prompts. This confirms the implementation dependency between the module and the retained long-term memory artifacts. It does \textbf{not}, by itself, establish that layered memory causes greater downstream behavioral or narrative diversity; that claim requires a matched comparison of agent outputs under enabled and disabled conditions.

The short M versus M+T comparison exposes a temporal-granularity trade-off---more activity and lower literal reuse, but more missing fields---rather than improved behavioral diversity, because opportunity counts differ and each condition has one unseeded run.

The two health-off runs (M+T enabled) did not complete the full 52-week horizon: the first terminated at week~21 due to a resume-path performance degradation, and the second at week~38 due to a WSL I/O failure. Because disabling the module also removes its health-specific warning and death mechanisms, and both runs are incomplete, we do not compare their health outcomes with the full system. A complete matched ablation remains for future work.

The all-off configuration of our reduced-scale port completed the planned 4-week baseline. Its short horizon is useful as an implementation reference but is insufficient for a longitudinal reliability comparison, so we do not use it to claim that the three adaptations collectively reduce failures.

\section{Conclusion}

On a single RTX 5070 Ti (12\,GB), our five-agent system completed 52, 52, and 50 weeks, totaling 154 system-weeks (770 agent-weeks), with no deaths or threshold-based health warnings.
The memory-off run tied L2/L3 artifacts to layered memory, while the short time toggle exposed a trade-off among activity volume, lexical reuse, and missing fields.
These observations do not establish causal behavioral effects, and the health module's contribution remains unresolved.

\section*{Limitations}

Several limitations warrant discussion.
First, our experiments use a single model (Qwen3-8B-AWQ) for both agent roles and God evaluations; a multi-model setup might yield richer agent differentiation.
Second, the concurrency limit serializes inference requests; at the tested five-agent scale, a 52-week run took roughly 33--40 hours. We did not measure scaling beyond this setting.
Third, we evaluate only the apartment world setting; generalization to other Agentopia worlds (school, workplace) remains untested.
The physical- and mental-health variables are simulator-internal state variables; they are non-clinical and have not been validated against medical or psychometric constructs.
Fourth, limited compute prevented a matched 52-week baseline and exhaustive ablations; additionally, we did not fix a global run-level random seed.
Fifth, our time-toggle comparison covers only 10 weeks, with one unseeded run per condition and health disabled; it is not a 52-week M+H leave-one-out ablation, so the time module's causal contribution to behavioral diversity remains unquantified.
Sixth, our narrative evidence consists of stored-memory examples and counts rather than an automated or human-rated measure of downstream behavioral diversity.
Seventh, the unflagged solo-output audit identifies lexical reuse but does not determine whether repeated wording corresponds to repeated actions, intentions, or social outcomes.
Finally, the \texttt{NO\_RESPONSE} statistic is defined over finalized activity records, not individual model calls, and therefore characterizes missing activity fields rather than serving as a direct inference-server failure rate.
The public research artifact at \url{https://github.com/luo675/Agentopia/tree/paper} is maintained as a fork of the upstream Agentopia repository and contains our implementation changes, validated configurations, derived audits, analysis scripts, and aggregate figure data.
It excludes raw run archives, initial persona data, model weights, and credentials; the upstream project and its attribution remain linked through the fork relationship.

% 审稿版（review）铁律：致谢必须留空或删掉，以免暴露身份！所以先注释掉，录用改 final 后再开。
%\section*{Acknowledgments}
%Your acknowledgments go here.

% ===== 参考文献 =====
% 这行告诉编译器：去读 custom.bib 生成参考文献列表。现在正文没引用，列表为空是正常的。
\bibliography{custom}

@article{wang2026agentopia,
  author    = {Wang, Xintao and Zheng, Sirui and Wu, Hongqiu and Li, Weiyuan and Huang, Jen-tse and Zhu, Minghao and Zu, Can and Deng, Qi and Wang, Jiawei and He, Qianyu and Wang, Heng and Wu, Xiaojian and Tao, Yunzhe},
  title     = {Agentopia: Long-Term Life Simulation and Learning in Agent Societies},
  journal   = {arXiv preprint arXiv:2606.07513},
  year      = {2026},
  doi       = {10.48550/arXiv.2606.07513},
  url       = {https://arxiv.org/abs/2606.07513},
}

@article{yang2025qwen3,
  author    = {Yang, An and others},
  title     = {{Qwen3} Technical Report},
  journal   = {arXiv preprint arXiv:2505.09388},
  year      = {2025},
  doi       = {10.48550/arXiv.2505.09388},
  url       = {https://arxiv.org/abs/2505.09388},
}

@inproceedings{kwon2023vllm,
  author    = {Kwon, Woosuk and Li, Zhuohan and Zhuang, Siyuan and Sheng, Ying and Zheng, Lianmin and Yu, Cody Hao and Gonzalez, Joseph E. and Zhang, Hao and Stoica, Ion},
  title     = {Efficient Memory Management for Large Language Model Serving with {PagedAttention}},
  booktitle = {Proceedings of the 29th Symposium on Operating Systems Principles (SOSP)},
  year      = {2023},
  pages     = {611--626},
  doi       = {10.1145/3600006.3613165},
  url       = {https://doi.org/10.1145/3600006.3613165},
}

@inproceedings{park2023generative,
  author    = {Park, Joon Sung and O'Brien, Joseph C. and Cai, Carrie J. and Morris, Meredith Ringel and Liang, Percy and Bernstein, Michael S.},
  title     = {Generative Agents: Interactive Simulacra of Human Behavior},
  booktitle = {Proceedings of the 36th Annual ACM Symposium on User Interface Software and Technology},
  year      = {2023},
  pages     = {1--22},
  doi       = {10.1145/3586183.3606763},
  url       = {https://doi.org/10.1145/3586183.3606763},
}

@inproceedings{zhou2024sotopia,
  author    = {Zhou, Xuhui and Zhu, Hao and Mathur, Leena and Zhang, Ruohong and Yu, Haofei and Qi, Zhengyang and Morency, Louis-Philippe and Bisk, Yonatan and Fried, Daniel and Neubig, Graham and Sap, Maarten},
  title     = {{SOTOPIA}: Interactive Evaluation for Social Intelligence in Language Agents},
  booktitle = {International Conference on Learning Representations},
  year      = {2024},
  url       = {https://proceedings.iclr.cc/paper_files/paper/2024/hash/b3075b88e583a0e98d8b24338a613060-Abstract-Conference.html},
}

@inproceedings{zhang2025agentsociety,
  author    = {Zhang, Jun and Yan, Yuwei and Yan, Junbo and Zheng, Zhiheng and Piao, Jinghua and Jin, Depeng and Li, Yong},
  title     = {A Parallelized Framework for Simulating Large-Scale {LLM} Agents with Realistic Environments and Interactions},
  booktitle = {Proceedings of the 63rd Annual Meeting of the Association for Computational Linguistics (Volume 6: Industry Track)},
  year      = {2025},
  pages     = {1339--1349},
  doi       = {10.18653/v1/2025.acl-industry.94},
  url       = {https://aclanthology.org/2025.acl-industry.94/},
}

@inproceedings{tang2025gensim,
  author    = {Tang, Jiakai and Gao, Heyang and Pan, Xuchen and Wang, Lei and Tan, Haoran and Gao, Dawei and Chen, Yushuo and Chen, Xu and Lin, Yankai and Li, Yaliang and Ding, Bolin and Zhou, Jingren and Wang, Jun and Wen, Ji-Rong},
  title     = {{GenSim}: A General Social Simulation Platform with Large Language Model based Agents},
  booktitle = {Proceedings of the 2025 Conference of the Nations of the Americas Chapter of the Association for Computational Linguistics: Human Language Technologies (System Demonstrations)},
  year      = {2025},
  pages     = {143--150},
  doi       = {10.18653/v1/2025.naacl-demo.15},
  url       = {https://aclanthology.org/2025.naacl-demo.15/},
}

@inproceedings{aher2023simulate,
  author    = {Aher, Gati V. and Arriaga, Rosa I. and Kalai, Adam Tauman},
  title     = {Using Large Language Models to Simulate Multiple Humans and Replicate Human Subject Studies},
  booktitle = {Proceedings of the 40th International Conference on Machine Learning},
  series    = {Proceedings of Machine Learning Research},
  volume    = {202},
  pages     = {337--371},
  year      = {2023},
  publisher = {PMLR},
  url       = {https://proceedings.mlr.press/v202/aher23a.html},
}

@article{argyle2023outofone,
  author    = {Argyle, Lisa P. and Busby, Ethan C. and Fulda, Nancy and Gubler, Joshua and Rytting, Christopher and Wingate, David},
  title     = {Out of One, Many: Using Language Models to Simulate Human Samples},
  journal   = {Political Analysis},
  volume    = {31},
  number    = {3},
  pages     = {337--351},
  year      = {2023},
  doi       = {10.1017/pan.2023.2},
  url       = {https://doi.org/10.1017/pan.2023.2},
}

@inproceedings{zhou2024reallife,
  author    = {Zhou, Xuhui and Su, Zhe and Eisape, Tiwalayo and Kim, Hyunwoo and Sap, Maarten},
  title     = {Is this the real life? Is this just fantasy? The Misleading Success of Simulating Social Interactions With {LLM}s},
  booktitle = {Proceedings of the 2024 Conference on Empirical Methods in Natural Language Processing},
  year      = {2024},
  pages     = {21692--21714},
  doi       = {10.18653/v1/2024.emnlp-main.1208},
  url       = {https://aclanthology.org/2024.emnlp-main.1208/},
}

@article{gao2024survey,
  author    = {Gao, Chen and Lan, Xiaochong and Li, Nian and Yuan, Yuan and Ding, Jingtao and Zhou, Zhilun and Xu, Fengli and Li, Yong},
  title     = {Large language models empowered agent-based modeling and simulation: a survey and perspectives},
  journal   = {Humanities and Social Sciences Communications},
  volume    = {11},
  pages     = {1259},
  year      = {2024},
  doi       = {10.1057/s41599-024-03611-3},
  url       = {https://doi.org/10.1057/s41599-024-03611-3},
}

@article{zeng2026toohuman,
  author    = {Zeng, Yongchao and Brown, Calum and Rounsevell, Mark},
  title     = {Too human to model: the uncanny valley of large language models in simulating human systems},
  journal   = {npj Complexity},
  volume    = {3},
  pages     = {13},
  year      = {2026},
  doi       = {10.1038/s44260-026-00075-1},
  url       = {https://doi.org/10.1038/s44260-026-00075-1},
}

@misc{star2026auditing,
  author       = {Star, Michelle and Aquilina, Andrew and Lin, Yu-Ru},
  title        = {Auditing Support Strategies in {LLM}s Through Grounded Multi-Turn Social Simulation},
  howpublished = {SocialLLM Workshop at ICWSM 2026},
  year         = {2026},
  url          = {https://openreview.net/forum?id=OX8jmYlFD4},
}

@misc{lazzaroni2026calibrated,
  author       = {Lazzaroni, Ruggero Marino and Prattes, Lorenz and Lasser, Jana},
  title        = {Calibrated but Autonomous: Inference-Time Bayesian Logit Correction for {LLM} Social Simulations},
  howpublished = {SocialLLM Workshop at ICWSM 2026},
  year         = {2026},
  url          = {https://openreview.net/pdf?id=9EZgvMOHXn},
}

@article{goel2025lifelong,
  author    = {Goel, Hitesh and Zhu, Hao},
  title     = {{LIFELONG SOTOPIA}: Evaluating Social Intelligence of Language Agents Over Lifelong Social Interactions},
  journal   = {arXiv preprint arXiv:2506.12666},
  year      = {2025},
  doi       = {10.48550/arXiv.2506.12666},
  url       = {https://arxiv.org/abs/2506.12666},
}

@inproceedings{xiong2026memory,
  author    = {Xiong, Zidi and Lin, Yuping and Xie, Wenya and He, Pengfei and Liu, Zirui and Tang, Jiliang and Lakkaraju, Himabindu and Xiang, Zhen},
  title     = {How Memory Management Impacts {LLM} Agents: An Empirical Study of Experience-Following Behavior},
  booktitle = {Proceedings of the 64th Annual Meeting of the Association for Computational Linguistics (Volume 1: Long Papers)},
  year      = {2026},
  pages     = {623--645},
  doi       = {10.18653/v1/2026.acl-long.27},
  url       = {https://aclanthology.org/2026.acl-long.27/},
}

@inproceedings{zhang2026lightmem,
  author    = {Zhang, Jiaquan and Zhang, Chaoning and Chen, Shuxu and Huang, Zhenzhen and Zheng, Pengcheng and Wang, Zhicheng and Guo, Ping and Mo, Fan and Bae, Sung-Ho and Zou, Jie and Wei, Jiwei and Yang, Yang},
  title     = {Lightweight {LLM} Agent Memory with Small Language Models},
  booktitle = {Proceedings of the 64th Annual Meeting of the Association for Computational Linguistics (Volume 1: Long Papers)},
  year      = {2026},
  pages     = {12914--12929},
  doi       = {10.18653/v1/2026.acl-long.588},
  url       = {https://aclanthology.org/2026.acl-long.588/},
}

@inproceedings{lin2024awq,
  author    = {Lin, Ji and Tang, Jiaming and Tang, Haotian and Yang, Shang and Chen, Wei-Ming and Wang, Wei-Chen and Xiao, Guangxuan and Dang, Xingyu and Gan, Chuang and Han, Song},
  title     = {{AWQ}: Activation-aware Weight Quantization for On-Device {LLM} Compression and Acceleration},
  booktitle = {Proceedings of Machine Learning and Systems},
  volume    = {6},
  year      = {2024},
  url       = {https://proceedings.mlsys.org/paper_files/paper/2024/hash/42a452cbafa9dd64e9ba4aa95cc1ef21-Abstract-Conference.html},
}

\end{document}